\documentclass[11pt, 
showpacs,preprintnumbers,amsmath,amssymb,aps]{revtex4}
\usepackage{amssymb}
\usepackage{amsmath,amssymb,graphicx}
\usepackage{graphicx}% Include figure files
\usepackage{dcolumn}% Align table columns on decimal point
\usepackage{bm}% bold math

\begin{document}

\title{Violation of CP and T in semileptonic decays due to scalar interactions}

\author{G. L\'opez Castro}
\email{glopez@fis.cinvestav.mx}
\affiliation{Departamento de F\'isica, Cinvestav, Apartado Postal 14-740, M\'exico 07000 D.F. M\'exico}
\author{L. L\'opez-Lozano}
\email{lao-tse@sirio.ifuap.buap.mx}
\affiliation{Instituto de F\'isica, Benem\'erita Universidad Aut\'onoma de Puebla, Apartado Postal J-48, C.P. 72570 Puebla, Pue., M\'exico}
\author{A. Rosado}
\email{rosado@sirio.ifuap.buap.mx}
\affiliation{Instituto de F\'isica, Benem\'erita Universidad Aut\'onoma de Puebla, Apartado Postal J-48, C.P. 72570 Puebla, Pue., M\'exico}

\begin{abstract}
Observing charge-parity (CP) or time-reversal (T) violations in the leptonic 
sector will give useful information to elucidate the nature of neutrinos.
CP-violating couplings in charged leptonic currents carry out the weak phases necessary to break these symmetries. Here we study the interference of $W^{\pm}$ and $H^{\pm}$ bosons mediated amplitudes as the origin of possible CP and T violation in semileptonic decays of $K^{\pm}$ mesons and $\tau^{\pm}$ leptons. We use the experimental bound on the T-odd transverse polarization asymmetry in $K^+_{\mu 3}$ decays to predict an upper limit on the CP violating effects in $\tau \to K \pi \nu$ decays. In the framework of this model with scalar-mediated interactions, we find that current limits on the former process indicate that the CP violating  effects in the latter are  much smaller than the limits reported so far.
\end{abstract}

\maketitle

\section{Introduction}

  The magnitudes of charge-parity (CP) and time-reversal (T) violations are intimately connected in quantum field theories of particle physics, which naturally satisfy CPT invariance. The violation of CP symmetry has been observed 
in decays of $K$ and $B$  mesons~\cite{Amsler:2008zzb}. The amounts of 
CP violation measured in these systems confirm that the 
mixing of quarks \cite{ckm} is the dominant mechanism responsible for 
these matter-antimater asymmetries in the hadronic sector. 
Since the standard model (SM) of elementary particles is a CPT 
conserving theory, CP violation entails a violation of T symmetry of 
similar magnitude in these systems . This was confirmed to be the case 
with measurements of the rate asymmetry in $K^0 \leftrightarrow 
\overline{K}^0$  conversion \cite{Angelopoulos:1998dv} and of the T-odd  
angular correlation between the planes of pion  and lepton pairs in  $K_L \to 
\pi^+\pi^-e^+e^-$ 
decay~\cite{AlaviHarati:1999ff}.

In analogy with the quark sector, the discovery that neutrinos are 
massive and mixed particles~\cite{neutrinos} makes possible the 
existence of CP and T violation in the leptonic sector.  Thus, the search for 
CP and T violations in leptonic currents are among the most important issues in experimental particle physics since they would provide important clues to establish the mechanism that generates neutrino masses and mixings.

  In this paper we explore the consequences of complex phases in the  Yukawa interactions of charged Higgs bosons of the Two Higgs Doublet model Type III (THDM-III)  \cite{thdmIII} for some semileptonic decays.  As is well known, two Higgs doublet models are classified as type I, II, III and IV according to the choice Yukawa couplings of the Higgs doublets \cite{thdmIII}. Among these models, the THDM-II is particularly attractive mainly because it coincides with the Higgs sector of the Minimal Supersymmetric Standard Model (MSSM), where each Higgs doublet couples separately to up- or down-type fermions at the tree-level. However, when radiative corrections are taken into account, the Higgs sector of the MSSM corresponds to the most general version (THDM-III) \cite{Kanemura:2009mk,Babu-Kolda} where the two Higgs doublets couple to both up- and down-type fermions. Thus, our main motivation for considering the THDM-III is that it bears a generic description of physics at a higher energy scale whose low energy imprints are the structure of Yukawa couplings.
 
We consider the effects of complex phases of THDM-III Yukawa couplings in the tranverse (T-odd) polarization asymmetry of the muon in $K^+ \to \pi^0 \mu^+ \nu$  
and in the CP asymmetry in   $\tau^{\pm}\to  K^{\pm}\pi^0 \nu$ decays. 
It is  worth to mention that the  SM contributions  to these decays are 
negligibly small \cite{Delepine:2005tw, bigisandabook} which makes the study of these decays a sensitive 
probe of New Physics phases.  Experimental searches for the polarization  
asymmetry in $K^+$ decays  have been reported by the KEK-E246 
Collaboration~\cite{Abe:2004px}. Possible signals of CP-violating effects induced by scalar charged boson exchange in $\tau$ lepton decays have been studied by the CLEO \cite{Bonvicini:2001xz}  and Belle \cite{bellecp} Collaborations. By  assuming  CPT invariance, we relate 
both asymmetries and show that the  current bound on the former sets very
stringent bounds on the latter asymmetry.

\section{T-odd polarization asymmetry in $K^+_{\mu 3}$ decay}
In this section we consider the transverse polarization asymmetry of 
 muons in  $K^+\to  \pi^0 \mu^+\nu_\mu$ decays. This T-odd observable 
is thought  to be particularly sensitive to new physics since the SM 
backgrounds (T-odd effects induced by final state interactions) are below the current experimental sensitivity.

Following \cite{Garisto1991} we define the transverse asymmetry as:
\begin{equation}
 \mathcal{P}^\perp_{3\mu}=\frac{|\mathcal{M}_T^+|^2-
|\mathcal{M}_T^-|^2}{|\mathcal{M}_T^+|^2+|\mathcal{M}_T^-|^2}
\end{equation}
where $\mathcal{M}_T^+$ (respectively  $\mathcal{M}_T^-$) denotes the 
amplitude when muons are emitted with spin component in 
the upward (downward) direction with respect to the decay plane in 
the $K^+$ rest frame.  The $\mathcal{P}^\perp_{3\mu}$ 
asymmetry vanishes in the SM at the tree-level, and receives a very 
small (of $O(10^{-5})$) contribution \cite{Efrosinin:2000yv} 
from  the  absorptive pieces of two photon loops intermediate 
states. The most stringent upper 
bound set by the KEK-E246 collaboration indicates 
$|\mathcal{P}^\perp_{3\mu}|\leq 0.0050$ at the 90$\%$ C.L  
\cite{Abe:2004px}.
Therefore, an observation  of the $\mathcal{P}^\perp_{3\mu}$ at the level of $\sim 
O(10^{-4})$  would indicate signals of New Physics (NP) beyond the SM. 

In the framework of the THDM-III extension of the SM, the two diagrams that contribute to the $K^+(p_K) \to \pi^0(p_{\pi})\mu(p,z) \nu(p')$ decay  are 
shown in Figure 1. The decay amplitude can thus be written as:
\begin{equation}
 \mathcal{M}= \mathcal{M}_{SM} +  \mathcal{M}_{NP}\ .
\end{equation}

The SM contribution is given by:
\begin{equation}
 \mathcal{M}_{SM}=\frac{G_F}{\sqrt{2}}V_{us}\overline{v}(p,z)\gamma_\mu
(1-\gamma_5)u(p')\langle \pi^0|\overline{u}\gamma^\mu 
s|K^+\rangle,
\end{equation}
where $V_{us}$ is the Cabibbo-Kobayashi-Maskawa  matrix element and $z$ is the muon 
polarization four-vector such that $z^2=-1$ and $p\cdot z=0$. 

The hadronic matrix element of the vector current can be expressed in terms 
of two form factors:
{\begin{equation}\label{factoresforma}
 \langle \pi^0|\overline{u}\gamma^\mu 
s|K^+\rangle=\frac{1}{\sqrt{2}}\lbrack (p_K+p_\pi)^\mu
f_+(q^2)+(p_K-p_\pi)^\mu f_-(q^2)\rbrack,
\end{equation}
where $q^2=(p_K-p_\pi)^2$ is the square of the momentum transfer, with 
$m_\mu^2 \leq q^2 \leq (m_K-m_\pi)^2$. At the tree-level in the SM the 
form factors are  real functions of $q^2$; a very small relative phase 
between both form factors can be induced by two photon intermediate 
states as discussed in \cite{Efrosinin:2000yv}.
\begin{figure}
  \includegraphics[width=10cm]{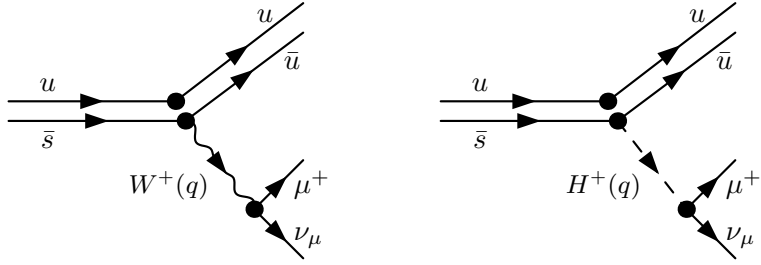}\\
  \caption{Feynman graphs for (a) SM and, (b) charged Higgs boson contributions to $K_{\mu 3}^+$ decay}\label{fig1}
\end{figure}

The exchange of a charged Higgs boson in Figure 1b gives rise to the 
following New Physics contribution 
\begin{equation}
\mathcal{M}_{NP}=\frac{G_F}{\sqrt{2}}N^L_{\mu\nu}(N^d_{us}+N^u_{us}) 
\overline{v}(p,z)(1 -\gamma_5)u(p')\langle 
\pi^0|\overline{u}s|K^+\rangle \ , \label{scalar-me}
\end{equation}
 with the following definition of the hadronic matrix element:
\begin{equation}
\langle \pi^0|\overline{u}s|K^+\rangle = \frac{1}{\sqrt{2}} f_H(q^2)\ .
\end{equation}
Taking the divergence of the vector current in Eq. (4) one gets
\begin{eqnarray}
(m_s-m_u)\langle \pi^0 | \overline{u}s| K^+ \rangle &=& \frac{1}{\sqrt{2}} 
\left[(m_K^2-m_\pi^2)f_+(q^2)+q^2f_-(q^2) \right]\nonumber \ ,\\ 
&\equiv &\frac{(m_K^2-m_{\pi}^2)}{\sqrt{2}}f_S(q^2)\ 
\end{eqnarray} 
where $m_{u,s}$ denote the current quark masses and we have introduced the scalar form factor $f_S(q^2)$ such that:
\begin{eqnarray}
f_S(q^2)&=& f_+(q^2)+\frac{t}{m_K^2-m_{\pi}^2}f_-(q^2)\ , \\
f_H(q^2)&=& \frac{(m_K^2-m_{\pi}^2)}{(m_s-m_u)} f_S(q^2)\ ,
\end{eqnarray}
with the normalization conditions $f_S(0)=f_+(0)=1$ and $f_H(0)=(m_K^2-m_{\pi}^2)/(m_s-m_u)$. In order to compare with experimental values, in this section we use the set $(f_+,\ f_-)$, while in the the case of $\tau^{\pm}$ decays (see next section) it becomes convenient to use $(f_+\ ,f_S)$.

The scalar couplings that appear in the amplitude (\ref{scalar-me}) are given by 
(for details about the assumptions made in deriving these couplings in the framework of the THDM-III, see \cite{thdm-yukawa,Cheng:1987rs}):
\begin{align}\label{couplings}
 N^d_{ji}&=\frac{2\sqrt{2}M_W}{gM_H}\sum_{k=1}^3 V_{jk}\left( \frac{ 
-\tan\beta}{v}M_d+\sec\beta\widetilde{Y}_2^{d\dagger}\right)_{ki},\\
N^u_{ji}&=\frac{2\sqrt{2}M_W}{gM_H}\sum_{k=1}^3 \left( \frac{-\tan\beta}{v}M_u 
+\sec\beta\widetilde{Y}_2^{u}\right)_{jk}V_{ki},\\
N^L_{l\nu}&=\frac{2\sqrt{2}M_W}{gM_H} \left( \frac{-\tan\beta}{v}M_L+ 
\sec\beta\widetilde{Y}_2^{L\dagger}\right)_{l\nu}\ ,
\end{align}
where $v=(\sqrt{2}G_F)^{-1/2}$, $\tan \beta=v_2/v_1$  is the ratio of vacuum expectation values of the two Higgs doublets and $M_{u,d,L}$ are 
diagonal mass matrices. 

The Yukawa couplings $\widetilde{Y}_2^{u,d,L}$ 
are non-diagonal complex matrices which will provide additional phases 
that induce CP and T violation both in the leptonic and quark sectors. 
Since the observed CP violation in the hadronic sector is well accounted 
by the CKM  mechanism \cite{ckm} we can make the simplifying assumption 
that the  phases in $\widetilde{Y}_2^{u,d}$ vanish, althought this will not be necessary in our analysis. Furthermore,  we note that the 
THDM-III itself does not impose any restriction on the 
$\widetilde{Y}_2^{u,d,L}$ couplings. One interesting possibility is that 
these Yukawa couplings behave like 
$(\widetilde{Y}_2^r)_{ji}=\sqrt{m_{j}m_{i}}\widetilde{\chi}^r_{ji}$ 
\cite{Cheng:1987rs,Papaqui-Rosado2}, where $\widetilde{\chi}_{ij}$ are dimensionless  complex parameters. 
Since lepton universality is not expected to hold for Yukawa 
interactions of charged Higgs boson, we can use this particular choice in 
later discussions.

  Let us return to our expression for the total decay amplitude of $K^+_{\mu 3}$ from Eqs. (2)--(5). If we use the Dirac equation for the induced scalar term of the SM contribution, we get:
\begin{equation}
 \mathcal{M}=\frac{G_F}{2}V_{us}\left\{f_+(q^2) (p_K+p_{\pi})^{\mu}\overline{v}(p,z)\gamma_\mu
(1-\gamma_5)u(p') + f_-^{\rm eff}(q^2)m_{\mu}\overline{v}(p,z)(1-\gamma_5)u(p')\right\}\ , 
\end{equation}
where we have defined the effective form factor:
\begin{equation}
f_-^{\rm eff}(q^2) \equiv f_-(q^2)+\frac{N_{\mu\nu}^L(N^u+N^d)_{us}}{V_{us}m_{\mu}} f_H(q^2)\ .
\end{equation}

  According to the calculations of Refs. \cite{Garisto1991}, a T-odd transverse polarization asymmetry in $K^+_{\mu 3}$ decays can be induced if there is 
a relative weak phase between the effective scalar form factor $f_-^{\rm eff}(q^2)$ and the vector form factor $f_+(q^2)$. In other words, if Im$(\xi)\not = 0$, where $\xi\equiv f_-^{\rm eff}/f_+$. Searches for a transverse polarization asymmetry in $K_{\mu 3}^+$ carried out by the KEK-E246 Collaboration have produced the upper bound $|{\rm Im}(\xi)| \leq 0.016$ at the 90$\%$ c.l \cite{Abe:2004px}. Using this bound, we get:
\begin{equation}
\left| \mbox{\rm Im}\left[ \frac{N_{\mu\nu_{\mu}}^L(N^d+N^u)_{us}}{V_{us}}\right] \right| \leq 
7.5 \times 10^{-4} \ , \label{bound-mu}
\end{equation}
where we have used $m_s-m_u \simeq 100$ MeV.
 Note that we are using the fact that the SM contribution to Im($\xi$), first term in eq. (14), is negligible with respect to current experimental bound. Note also that, in the general case, only the product of leptonic and quark Yukawa couplings can be constrained from this observable.

\section{CP asymmetry in $\tau$ lepton decays}

  Recently, the possibility to observe a CP asymmetry in $\tau$ lepton 
decays has been discussed by several authors 
\cite{cptau,Bigi:2005ts,Delepine:2005tw}. A `known' 
CP 
rate  asymmetry is expected to occur in $\tau^{\pm} \to 
K_{S,L}\pi^{\pm}\nu$ \cite{Bigi:2005ts} 
 with a magnitude similar to the one measured in semileptonic 
$K_L$ decays. On the other hand, a CP asymmetry in $\tau^{\pm} \to K^{\pm} 
\pi^0\nu$ becomes 
interesting as far as the SM prediction in this case turns out 
to be negligibly small \cite{Delepine:2005tw}. So far, signals of CP 
violation that are induced by scalar interactions in decays of the $\tau$ 
lepton into kaons  have been investigated by the CLEO collaboration \cite{Bonvicini:2001xz} by looking at the ratio of CP-odd and  CP-even terms in the square of the matrix element.

In this paper we focus on the CP asymmetry in $\tau^{\pm} \to 
K^{\pm}\pi^0\nu$ decays. In the THDM-III, the Feynman diagrams 
that contribute to these decays are shown in Figure 
2. Since the above $\tau$ lepton decay is related to $K^+_{\mu 3}$ by 
crossing symmetry and by replacing $\mu \to \tau$, we can write the 
SM  amplitude as follows:
\begin{figure}
  \includegraphics[width=10cm]{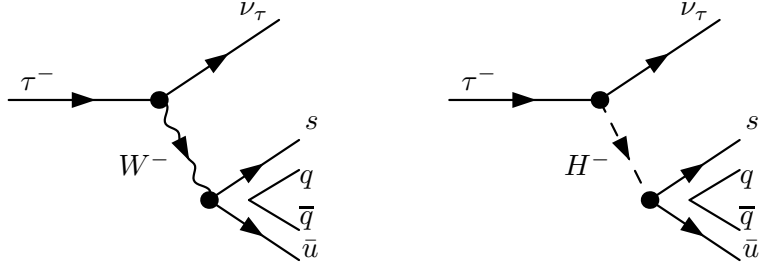}\\
  \caption{Same as in Figure 1 for $\tau^- \to K\pi\nu$ decay}\label{fig2}
\end{figure}

\begin{equation}
\mathcal{M}_{SM}
= \frac{G_F}{2}V_{us}\overline{u}(p')\gamma^\mu (1-\gamma_5)u(p) 
   \times \left[ (p_K-p_{\pi})_{\mu} f_+(q^2) 
+(p_K+p_{\pi})_{\mu}f_-(q^2) \right] \ , \label{me}
\end{equation}
where $q=p_K+p_{\pi}$ and $(m_K+m_{\pi})^2 \leq q^2 \leq m_{\tau}^2$. 
The kinematical conditions in this decay allows the form factors to 
acquire imaginary parts (strong CP phases) due to the production of 
strange resonances decaying into $K\pi$. Since the exchange of a charged Higgs boson in the THDM-III, giving rise to the amplitude  
\begin{equation}
{\cal M}_{NP}=\frac{G_FV_{us}}{\sqrt{2}} \left(N^u+N^d\right)_{us} \bar{u}(p')(1+\gamma_5)u(p)\langle K\pi|\bar{u}s |0\rangle \ ,
\end{equation}
contributes only to the induced scalar form factor $f_-(q^2)$,  a 
non-vanishing relative weak phase between between both form 
factors is induced. Thus, the conditions to generate a CP asymmetry between 
the 
decays of positive and negative $\tau$ leptons are fulfilled. 

  In order to compare our expressions with the experimental bounds reported in \cite{Bonvicini:2001xz}, we replace the {\it induced} scalar form factor in Eq. (\ref{me}) by the {\it true} scalar form factor $f_S(q^2)$ defined in Eq. (8). 
The form factors $f_+(q^2)$ and $f_S(q^2)$ are dominated, respectively, by the 
production of the $K^*(892)$ and $K_0(1430)$ resonances, which provides 
the required strong CP phases. 

  Thus,  up to overall weak couplings, the total decay amplitude $\mathcal{M}=\mathcal{M}_{SM}+\mathcal{M}_{NP}$  can be written as:
\begin{eqnarray}\label{ampform}
 \mathcal{M}\sim 
f_+(q^2)\left[(p_K-p_\pi)_{\mu}-\frac{m_K^2-m_\pi^2}{q^2} 
q_{\mu} 
\right]\overline{u}(p')\gamma^\mu(1-\gamma_5)u(p)+ f_S(q^2) M\Lambda
\overline{u}(p')(1+\gamma_5) u(p)   
\, 
\end{eqnarray}
where the effective coupling that constains the SM and NP contributions is:
\begin{equation}
M \Lambda=m_\tau\frac{m_K^2-m_{\pi}^2}{t}+\frac{N_{\tau\nu}(N^u+N^d)_{us}}{V_{us}} 
\frac{m_K^2-m_\pi^2}{(m_s-m_u)}\ ,
\end{equation}
and $M$ is a mass scale of order 1 GeV \cite{Bonvicini:2001xz}.

The unpolarized decay probability $\sum_{pol}|\mathcal{M}|^2$
contains CP-even and CP-odd terms \cite{Bonvicini:2001xz,bellecp}. By using as the optimal observable the ratio of the CP-odd to the  CP-even probabilities, the CLEO collaboration \cite{Bonvicini:2001xz} was able to derive the following constraint:   $-0.172 \leq$ Im($\Lambda$) $\leq 0.067$ at the 90$\%$ C.L. From Eq. (19), it follows the corresponding constraint for the charged Higgs couplings:
\begin{equation}
 -0.076 \leq \mbox{\rm Im}\left[ \frac{N_{\tau\nu_{\tau}}^L(N^d+N^u)_{us}}{V_{us}}\right]  \leq 0.030  \ . \label{bound-tau}
\end{equation}
This bound is weaker by at least a factor of 40 than the one derived in Eq. (\ref{bound-mu}), althought both refer to different leptonic vertices.

\section{Comments and conclusions}

   If we assume the universality of leptonic Yukawa couplings ($N_{\tau\nu_{\tau}}=N_{\mu\nu_{\mu}}$), from Eqs. (\ref{bound-mu}) and (19) we get $|{\rm Im}(\Lambda)|\leq 0.0017$ for the CP violating phases in tau decays; this bound is stronger than the limit obtained from direct searches \cite{Bonvicini:2001xz}. Of course, universality is a strong assumption since we know that Higgs boson interactions distinguish among different generations. We can thus assume, as in the case of the models of Refs. \cite{Cheng:1987rs,Papaqui-Rosado2}, that the leptonic couplings behave as $N_{\tau\nu_\tau}\simeq\sqrt{\frac{m_\tau}{m_\mu}}N_{\mu\nu_\mu}$. In this case Eqs. (\ref{bound-mu}) and (19) give $|\text{Im}(\Lambda)|\leq 0.0069$, which is still one order of magnitude stronger than the bound obtained from direct searches in $\tau$ decays by the CLEO Collaboration \cite{Bonvicini:2001xz}. 

  Thus, if CP violation arises solely from scalar interactions, we can conclude that current bounds on T-odd transverse muon polarization in $K_{\mu 3}^+$ decays implies a bound on CP violation in  $\tau \to K\pi\nu_{\tau}$ that is stronger than direct searches. A proposal to improve the current limits on the transverse muon polarization in kaon decays by a factor of 20$\sim$50 has been raised as a sequel of the KEK-E246 experiment (see for instance second of Refs. in \cite{Abe:2004px}). Such an improvement would reduce even more the possibilities to observe any signal of CP violation, as induced by scalar charged currents, in $\tau \to K \pi \nu$ decays \cite{Bonvicini:2001xz}. 

Semileptonic decays allow us to get bounds on the product of quark and lepton Yukawa couplings. It is interesting to note that if we assume that CP violation in the quark sector is due only to the CKM mechanism, the bounds derived in Eqs. (\ref{bound-mu}) and (\ref{bound-tau}) can be transformed into constraints on the CP-violating phases of lepton currents provided one uses constraints for the magnitudes of Higgs boson couplings from other processses(for instance, from $K^+_{\mu 2}$ and $\tau \to K\nu$ decays). Before closing this section let us mention that the CP-violating effects considered in this paper do not affect the CP-violating charge asymmetry in $K_L \to \pi^{\pm} l^{\mp}\nu_l$ decays.

\begin{acknowledgments}
The authors are grateful to Conacyt (M\'exico) for financial support.
\end{acknowledgments}

\end{document}